\documentclass[12pt]{article}
\pdfoutput=1
\usepackage{cite,graphicx,amsmath,amssymb}
\usepackage{graphicx}
\usepackage{multicol}
\usepackage{xcolor}

\newcommand{\be}{\begin{equation}}
\newcommand{\ee}{\end{equation}}
\newcommand{\bea}{\begin{eqnarray}}
\newcommand{\eea}{\end{eqnarray}}
\begin{document}

\thispagestyle{empty}
\vspace*{-.2cm}
\noindent

\vspace*{1.2cm}
\begin{center}

{\Large\bf The Weak Gravity Conjecture \\[.4cm]
and the Axionic Black Hole Paradox}
\\[2cm]

{\large Arthur Hebecker and Pablo Soler\\[6mm]}

{\it
Institute for Theoretical Physics, University of Heidelberg, 
Philosophenweg 19,\\ D-69120 Heidelberg, Germany\\[3mm]

{\small\tt (\,a.hebecker,\,p.soler~@thphys.uni-heidelberg.de\,)} }\\[.5cm]
\today
\\[1.6cm]

{\bf Abstract}
\end{center} 

In theories with a perturbatively massless 2-form (dual to an axion), a paradox may arise in the process of black hole evaporation. Schwarzschild black holes can support a non-trivial Wilson-line-type field, the integral of the 2-form around their horizon. After such an `axionic black hole' evaporates, the Wilson line must be supported by the corresponding 3-form field strength in the region formerly occupied by the black hole. In the limit of small axion decay-constant $f$, the energy required for this field configuration is too large. The natural resolution is through the presence of light strings, which allow the black hole to ``shed" its axionic hair sufficiently early. This gives rise to a new Weak-Gravity-type argument in the 2-form context: Small coupling, in this case $f$, enforces the presence of light strings or a low cutoff. We also discuss how this argument may be modified in situations where the weak coupling regime is achieved in the low-energy effective theory through an appropriate gauging of a model with a vector field and two 2-forms.

\newpage

\section{Introduction}
Very roughly speaking, the Weak Gravity Conjecture (WGC) says that a U(1) gauge theory with coupling $g\ll 1$ can only be consistently coupled to quantum gravity if there are charged states or even a cutoff at the scale $gM_P\ll M_P$ \cite{ArkaniHamed:2006dz}. This is expected to extend to $(p+1)$-form gauge theories with charged $p$-branes in any number of dimensions.\footnote{Notice that our convention for designating generalized gauge theories differs from that of, e.g.~\cite{Banks:2010zn,Gaiotto:2014kfa}. Here, standard theories of charged particles coupled to $1$-form gauge fields are refered to as $1$-form gauge theories, rather than $0$-form ones.} The simplest way to formulate the analogous statement about a low cutoff is then to say that $\Lambda^{p+1}\sim g$ in Planck units (see \cite{Banks:2006mm, Cheung:2014vva, Bellazzini:2015cra, Nakayama:2015hga, Heidenreich:2015nta, Harlow:2015lma, Ibanez:2015fcv, Hebecker:2015zss, Heidenreich:2016aqi, Montero:2016tif, Saraswat:2016eaz, Ooguri:2016pdq, Freivogel:2016qwc, Danielsson:2016mtx, Cottrell:2016bty, Banks:2016xpo,Hebecker:2017wsu} for a selection of recent related work).

Much of the recent phenomenological interest in the WGC derives from its potential power to constrain axion inflation \cite{delaFuente:2014aca, Rudelius:2015xta, Montero:2015ofa, Brown:2015iha, Bachlechner:2015qja, Hebecker:2015rya, Brown:2015lia, Heidenreich:2015wga, Junghans:2015hba, Kooner:2015rza, Kaloper:2015jcz, Kappl:2015esy, Choi:2015aem, Hebecker:2016dsw, Klaewer:2016kiy, Herraez:2016dxn,Dolan:2017vmn}: To this end, one views axion models as 0-form gauge theories in the regime where the axionic coupling $1/f$ is small (i.e. $f\gg M_P)$. However, fundamental justifications for the WGC both in the 1-form, the 0-form and various other cases and different regimes are difficult to obtain. Thus, we believe that supplying {\it any} arguments for or against it, even if in a slightly unusual setting, is important. 

It is our aim to analyse possible arguments in favor of the WGC on the dual side of the more familar axion$\,/\,$large-field-inflation case mentioned above. Indeed, we dualise the 0-form $\varphi$ to a 2-form, $dB_2=f^2*d\varphi$, and consider the action
\be
-\int d^4x\,\sqrt{-g}\,\frac{1}{f^2}|dB_2|^2+\int\limits_{\rm worldsheet} B_2\,.
\ee
We want to understand whether we can have small $f$ without light charged objects, i.e. without light strings. In the extreme, the tension $\sigma$ of the lightest string might be $\sigma \sim M_P^2$, but still $f\ll M_P$ (see~\cite{Junghans:2015hba} for constraints on such regimes in perturbative string compactifications). Is this inconsistent in any tangible way?

The standard magnetic WGC would have to argue about the smallest instanton not yet being a `black hole' (in this case wormhole or gravitational instanton). This is clearly questionable. The electric WGC would have to argue about the stability of extremal black strings. But such macroscopic objects appear rather pathological (having a deficit angle greater than $2\pi$) and, morevoer, possess no horizon.\footnote{
One 
might try to overcome this by demanding instead that two same-charge microscopic strings should always repel.
} 
What is worse, both routes rest simply on assumptions, such as `stable extremal black holes (BHs) must not exist' (see however \cite{Cottrell:2016bty} for recent progress towards justifying this requirement rigorously).

Here we suggest a completely different approach and, with certain caveats to be explained, arrive at an actual physical inconsistency for parametrically small $f$ in the absence of light strings. 

The basic idea is to consider axionic BHs as in \cite{Bowick:1988xh}. Such objects are distinguished from  standard Schwarzschild BHs exclusively by having non-zero values of $b\equiv \int B_2$, where the integral is over any sphere homotopic to their horizon. Such a `Wilson-loop variable' $b$ is locally unobservable and carries no energy (the field strength $H_3=dB_2$ vanishes), and so the axionic BH behaves at the semi-classical level exactly as a Schwarzschild one. 

Very roughly speaking, the problem arises because the BH is expected to shrink via Hawking radiation, maintaining the value of $b$, at least up to a radius $r$ of the order of the Planck length. While the later stages in the life of the BH cannot be properly described without a UV complete theory, it is clear that the only way in which a complete evaporation can be consistent with a non-zero $b$ is through a `leftover' field strength:
\be
\int_{V(r)} H_3=\int_{S^2(r)} B_2\sim b\sim{\cal O}(1)\,.
\ee
Here the sphere is the boundary of the volume $V(r)$, the latter being a ball of radius $r$ at the place where the BH used to be. But in the regime of small $f$, the prefactor $1/f^2$ of the field energy stored in $H_3$ is huge. It is in fact much larger than the available energy $\sim {\cal O}(M_P)$ of the smallest semiclassical BH which was still controlled in the low-energy effective theory. Unless there is a remnant, we have a contradiction.\footnote{Similar ideas have been used in~\cite{Dvali:2016mur, Dvali:2016sac} in a somewhat different context to argue that certain types of global symmetries (carried by skyrmions/baryons) can be preserved in BH evaporation.}

This contradiction is resolved if light strings are present. Strings `lassoing' the black hole do interact with the $B_2$ integral, and indeed generate an effective potential for the Wilson loop variable $b$. This is analogous to the familiar potential for the Wilson line variable $\oint A_1$ in a standard (1-form) gauge theory compactification from 5d to 4d. As long as the BH radius $R$ is much larger than the length scale set by the string tension, $R\gg 1/\sqrt{\sigma}$, the strings are irrelevant: The effective potential induced by a virtual string `lassoing' the BH has a suppression  $\sim\exp(-4\pi\sigma R^2)$ and is negligible. However, once $R\sim 1/\sqrt{\sigma}$, the effective potential for the `Wilson loop variable' $b$ ceases to be exponentially suppressed. Now $b$ is driven to zero dynamically near the BH, the value of $b$ at larger radii is supported by a non-vanishing field strength $H_3$ near the BH, and the BH can eventually disappear without a trace. The problem with the evaporation of axionic BHs is resolved if light strings exist.\footnote{The appearance of a potential due to lassoing strings is not undisputed \cite{Coleman:1991ku}. Indeed, in a partition function calculation in the scalar field basis, a single lassoing string has infinite action.  First, this does not preclude single-instanton contributions in the $B_2$ basis. Second, even staying in the scalar field basis, nothing speaks against contributions from even numbers of lassoing strings. This is analogous to calculating the familiar instanton-induced cosine potential in quantum mechanics, but using a basis of fixed discrete momenta, such that only even numbers of instantons occur. The toy model calculation of \cite{Hebecker:2016dsw} shows that a potential can still be derived in this way. Moreover, if it still turned out that a potential in the strict sense does not arise, tunneling processes corresponding to lassoing strings are certainly possible and can give rise to a dynamical effect on $b$. In any event we can argue, without directly referring to light strings, that the dynamical activation of $b$ kicks in at a scale $R\sim \Lambda^{-1}$ given by the cutoff of the theory. A bound on such a cutoff, analogous to the {\it magnetic} rather than {\it electric} WGC, would then be obtained. Although we mostly refer to light strings throughout this work, both points of view are related by associating the cutoff to the tension of charged objects, i.e. $\Lambda\sim \sqrt{\sigma}$.}

The rest of this paper is organized as follows. In section~\ref{sec:problems} we estimate the specific parametric bounds on the string tension that result from different evaporation rates of small black holes. We spell out in section~\ref{sec:caveats} several assumptions implicit in our arguments and the physical effects that motivate them. Finally, in section~\ref{sec:winding} we study a setup in which a massless 2-form field with small effective $f$ arises via alignment, upon Higgsing a linear combination of two $B_2$ fields, in analogy to~\cite{Hebecker:2015rya,Saraswat:2016eaz}. Our arguments suggest the appearance of light monopoles in such a theory.

The simultaneous work~\cite{Madrid} also discusses generalized (global) symmetries and their possible problems from a somewhat different perspective. 

\section{The final moments of an axionic black hole}\label{sec:problems}
As we have discussed in the Introduction, a contradiction arises in the process of evaporation of axionic black holes in the absence of light strings.\footnote{There are several important caveats to this simple reasoning that must be considered to reach our conclusions. In order to avoid obscuring our results, we postpone their discussion to later sections.} This implies a parametric upper bound on the tension of strings (for a given gauge coupling $f$). In order to derive the parametric form of the bound, we need to make certain assumptions about the latest stages of the evaporation of axionic BHs. We consider next two possibilities and derive the parametric form of the constraints that arise in each case.

\subsection{Immediate breakdown at critical radius}\label{sec:immediate}
The simplest assumption, or at least the assumption leading to the simplest estimate, is a catastrophic, explosion-like evaporation of the BH at the moment when it reaches the critical radius $R_c\equiv 1/\sqrt{\sigma}$. This is not a totally unnatural expectation: According to the arguments of \cite{Heidenreich:2015nta,Heidenreich:2016aqi,Montero:2016tif}, not just one but many string states should come in and an extreme growth of the number of degrees of freedom with energy may indeed lead to an instantaneous evaporation at a temperature $T_c\sim 1/R_c\sim \Lambda\equiv \sqrt{\sigma}$. 

In this case, there is no time for any effect arising from the dynamical strings to propagate. The $H_3$ field they dynamically induce is limited to a ball of radius $R_c$, the total resulting field energy being
\be 
E\sim \frac{1}{f^2}\int_{V(R_c)}d^3x\,\sqrt{g}\,|H_3|^2\sim \frac{b^2}{f^2R_c^3} \sim \frac{1}{f^2R_c^3}\,.
\ee
Now, we should require $E\lesssim M(R_c)\sim R_c M_p^2$ if energy is to be conserved in the final stage of the evaporation. This leads to a lower bound on the string tension $\sigma$ of the form
\be\label{eq:bound1}
\sigma\lesssim f \cdot M_p\,,\qquad\text{or}\qquad \Lambda^2\lesssim f\cdot M_p
\ee
This simple parametric estimate coincides precisely with the standard WGC bound for strings.

The origin of this bound is conceptually very simple, and perhaps finds a generalization to other setups (in particular to different dimensions and for objects of different co-dimension). On the one hand, a non-vanishing generalized Wilson-line can be supported by a topologically non-trivial $p$-cycle of a gravitational object with energy $\sim M_P^{d-2}$. On the other hand, in a trivial topology, a field strength $H_{p+1}$ that supports the same Wilson-line has an energy that scales with the coupling constant $g$ as $\sim 1/g^2$. When both situations are related by a dynamical change in the topology, energy conservation imposes a bound $M_P^{d-2}\gtrsim 1/g^2$. Of course, one needs to complete this inequality to make it dimensionally consistent. If one can argue that there is only one other scale $\Lambda\sim R^{-1}_c$ involved in the process, the WGC is recovered.  

\subsection{Slow evaporation and spread flux}\label{sec:slow}
While the result~\eqref{eq:bound1} is suggestive, the assumption of immediate BH evaporation at a critical temperature is likely too naive. We may obtain a much more conservative bound if we assume that, after the effective potential is activated at a temperature $T_c\sim 1/R_c\sim \Lambda\equiv\sqrt{\sigma}$, the $H_3$ flux induced has a time $t_{ev}$ to spread out radially at the speed of light before the BH completely evaporates. Assuming that in its latest stages the BH still radiates energy at rate $-dM/dt\sim M_p^4/M^2$, the evaporation time from the critical mass $M_c\sim R_c M_p^2$ to zero scales as\footnote{Given the assumptions we have to make about the latest stage of BH evaporation, our bounds on the tension of strings can only be considered estimates. In fact, the numerical factors we are neglecting can be very large. For example, the numerical factor in front of $t_{ev}$ is $5120\pi$.} $t_{ev}\sim M_c^3/M_p^4\sim M_p^2/\sigma^{3/2}\gg R_c$.
In this scenario, the left-over field $H_3$ that accounts for the Wilson loop variable $b$ after the BH disappears has spread out to a ball of radius $R_c+t_{ev}\approx t_{ev}\sim M_p^2/\sigma^{3/2}$. The energy stored in such a flux scales as
\be 
E\sim \frac{1}{f^2}\int_{V(t_{ev})}d^3x\,\sqrt{g}\,|H_3|^2\sim \frac{b^2}{f^2 t_{ev}^3} \sim \frac{\sigma^{9/2}}{f^2 M_p^6}\,.
\ee
Arguing as before that this energy should be less than the mass of the axionic BH of radius $R_c$, we obtain a bound on the string tension $\sigma$ of the form
\be\label{eq:bound2}
\sigma\sim\Lambda^2\lesssim f^{2/5} \cdot M_p^{8/5}\,.
\ee
This bound is much weaker than the one obtained in~\eqref{eq:bound1} (which coincides with the WGC). It would become even weaker if the BH decay process slowed down for some reason at the latest stages of the evaporation, and the $H_3$ flux had more time to spread over a broader region. In fact, the bound would disappear entirely if the decay never reached an end (in this case there would be a remnant which could by itself support a non-zero value of $b$). We find it reasonable to assume that the decay will not slow down. By contrast, we think it is likely that the right bound derived this way will lie somewhere between~\eqref{eq:bound1} and~\eqref{eq:bound2}.

\section{Infrared divergences and quantum effects}\label{sec:caveats}
We have so far glossed over some key issues that need to be addressed before accepting the conclusions of last section.

\subsection{Non-perturbative effects}

A first question one should ask when considering axionic BHs is whether the variable $b=\int B_2$ is measurable and has a physical meaning at all~\cite{Preskill:1990bm,Coleman:1991ku,Krauss:1990gx}. Since $B_2$ is locally pure gauge, it does not exist in a local sense. The way to measure it is via an Aharonov-Bohm (AB) type experiment, where strings lassoing the BH acquire a phase proportional to $b$ which can then be manifested in an interference pattern.  

A problem arises because the energy stored in an infinitely long string is logarithmically divergent in the IR.  Strings are hence confining, and there is a limit in the maximum radius $R_{max}\sim f^{-1}\exp(M_P^2/f^2)$ that a string loop can reach  \cite{Cohen:1988sg,Polchinski:2005bg,Banks:2010zn}. In order for the axionic hair on a BH to be measurable, we have to make sure that the AB experiment can be performed, i.e. that the BH's size is smaller than $R_{max}$. Given the exponential dependence of $R_{max}$ with respect to $f$ (recall that we are interested in the $f\ll M_P$ limit), this condition can be easily satisfied.

Another effect that we have not yet considered is that induced by instantons. When non-perturbative effects are taken into account, there is no strictly massless propagating degree of freedom (other than the graviton) in the theory. The instantons induce a mass gap, albeit an exponentially suppressed one.  Under certain conditions this can be described in terms of a coupling of $B_2$ to a non-dynamical 3-form (see~\cite{Dvali:2005an, Dvali:2005ws, Dvali:2013cpa,Dvali:2016uhn,Garcia-Valdecasas:2016voz}), but is most easily seen in the dual description where instantons generate a periodic potential for the axion.

Recall that, upon circumnavigating a string, the axion field shifts by a period $\phi \to \phi + 2\pi$. In the absence of a potential, this shift will be uniformly distributed around the string, i.e. $\phi\sim \theta$, where $\theta$ is the angle that parametrizes the winding trajectory. When a potential $V(\phi)$ is present, however, $\phi$ will tend to remain at its minimum for most of the trajectory. In this case, the string will be the boundary of a domain wall where the field $\phi$ jumps discretely. As before, this puts a limit in the maximum size $R_{max}$ of the string loops one can consider. Beyond $R_{max}$, strings loops are unstable due to nucleation of smaller string loops which eat up the axionic domain wall. Also as before, this effect is non-perturbatively suppressed (the axionic potential scales typically as $e^{-M_P/f}$) and can be safely ignored for sufficiently small strings. 

One may ask, still, whether axionic BHs can even exist once instantons are taken into account. One possibility would be that instanton effects induced a non-perturbative potiential to the Wilson-line variable $b$. In this case,  a non-zero $b$ could not spread all the way to infinity, since this would carry an infinite potential energy, independently of how small the instanton effects are. However, one may simply consider a pair of axionic BHs, with opposite value of $\int B_2$ so that the system still has a finite energy. As long as the non-perturbative potential induced is small enough, it will take an exponentially long time for the system to relax to its stable configuration (by merging the two axionic BHs). 
In this note we implicitly work in a regime of parameters where instanton effects are highly suppressed and do not play a role for the relatively small black holes that we are considering. 

It is  finally also interesting to describe the $B_2$ Wilson-line in the dual axionic picture. Consider for simplicity a quantum mechanical model, obtained by compactifying a 4d theory with a massless $B_2$ field on a $T^3$, parametrized by coordinates $(x,y,z)$. We can turn on a Wilson-line along any of the torus two-cycles, say  $T^2_{xy}$. The quantum mechanical canonical momentum of $B_{xy}$ is given by $\partial_t B_{xy}$. By Heisenberg's uncertainty principle, a state with highly localized value of the  Wilson line will correspond to a very broad superposition of states with fixed $\partial_t B_{xy}$. Upon 4d Hodge duality, $d\phi=\ast_4 B_2$, such states correspond to axion flux turned on along the one-cycle dual to $T^2_{xy}$, i.e. our original state with Wilson-line $B_{xy}$ will be described as a superposition of states with different values of $\partial_z \phi$. The value of the Wilson line corresponds to the phase, analogous to the $\theta$-angle, introduced in this superposition. In the case of an axionic BH, the two-cycle is the (homology class of the) horizon, and its dual corresponds to the (non-compact) radial direction. Hence, in the scalar language, a non-trivial value of the axionic hair $b$ corresponds to the phase in the superposition of states with non-trivial axionic gradient $\partial_r \phi$. Of course, once instanton effects are taken into account and a non-perturbative potential for $\phi$ is generated, the duality between $\phi$ and $B_2$ becomes more involved.

\subsection{Quantum vs. classical}
It is essential to understand under which conditions the Wilson line variable $b$ of a BH can be thought of as a classical degree of freedom. To this end, consider the action
\be
S\sim -\int d^4x\sqrt{-g}\frac{1}{f^2}|dB_2|^2
\ee
in the BH background
\be
ds^2=-(1-R/r)dt^2+(1-R/r)^{-1}dr^2+r^2d\Omega^2\,.
\ee
Assuming that $B_2$ is proportional to the normalized harmonic 2-form on $S^2$, parametrized by $b=b(t,r)$, gives
\be
S\sim \int dt \int_{R}^\infty\,\, \frac{dr}{r^2}\cdot\frac{1}{f^2}\,\left[(1-R/r)^{-1}\,(\partial_t b)^2-(1-R/r)\,(\partial_r b)^2\right]\,.
\ee
We see that the kinetic term diverges near the horizon. This agrees with the intuition that any dynamics near the horizon should be extremely slow in the time variable suitable for the observer at infinity. Furthermore, the gradient term goes to zero near the horizon. This is again intuitive since there should be no suppression of configurations where the values of $b$ very near the horizon and further away differ significantly. Indeed, the (effectively frozen) value of $b$ near the horizon should not be able to influence an observer at a certain distance. We thus take it for granted that, in total, the effect of the near-horizon region in the above action will be as follows: The BH horizon does not provide any boundary condition for $b$, in agreement with the intuitive expectation from the no-hair theorem. 

This can be modelled by excising a sphere of some radius $\gtrsim R$ (we do not care about ${\cal O}(1)$ factors) and imposing Neumann boundary conditions on $b$. At our qualitative level of analysis we can then also set $(1-R/r)$ to unity and simply write
\be
S\sim \int dt \int_{R}^\infty\,\, \frac{dr}{r^2}\cdot\frac{1}{f^2}\,\left[(\partial_t b)^2-(\partial_r b)^2\right]\,.
\ee
Here we took the lower limit of integration to be $R$ for simplicity, although it should of course be slightly larger as explained above.

Crucially, the $r$ integral now converges, such the quantum mechanical model for the zero mode (the $r$-independent mode of $b$) reads
\be
S\sim \int dt \,\frac{1}{f^2R}\,(\partial_t b)^2\,.
\ee
In other words, the dynamics is the same that one would obtain from a compactification to one dimension on a compact 3d space of typical radius $R$ with one non-trivial two-cycle. 

We are thus dealing with quantum mechanics of a variable with period $2\pi$ and a single mass scale $f^2R$ introduced through the kinetic term. Adopting textbook knowledge to our setting, it is clear that this mass scale translates into a time-scale which governs the spread of an optimally localized gaussian wave packet to the maximal width of $2\pi$. On time scales shorter than 
\be
t_{qm}\sim 1/(f^2R)
\ee
we can then think of our effective potential, induced by lassoing strings, as of a classical force acting on the classical variable $b$.

Now, in our `immediate breakdown' scenario, the typical time scale is $t_c\equiv R_c$. It is simply the time a signal neads to travel across the relevant region of space. For our classical analysis to be meaningful one should then require that this typical time is small enough:
\be
t_c<t_{qm}\sim 1/(f^2R_c)\,\,,\qquad \mbox{that is} \qquad \sigma>f^2\,.
\ee
This comfortably contains the range of very high string tensions constrained by the WGC and by our analysis.

For the `slow-evaporation' scenario, the much longer time scale of evaporation of a BH with radius $R_c$ is the relevant one. As we saw, this is $t_{ev}\sim M_p^2/\sigma^{3/2}$. We now have to impose that
\be
t_{ev}<t_{qm}\sim 1/(f^2R_c)\,\,,\qquad \mbox{that is} \qquad \sigma>f\cdot M_p\,.
\ee
This is precisely the opposite as the WGC bound, but still fits comfortably within the weaker bound derived in~\ref{sec:slow}. 

Crucially, finding this condition does not endanger our general logic. Indeed, if $\sigma<f M_P$, the WGC is satisfied and we have nothing to say. If, however, $\sigma>f M_P$, we are allowed to think classically about the dynamics of $b$ during the evaporation of BHs which are small enough for lassoing strings to be excited. Thus, the analysis of Sect.~\ref{sec:slow} is a posteriori justified.

\section{Systems with (aligned) multiple axions}\label{sec:winding}
A recurring question when addressing quantum gravity constraints on axion decay constants, is how these extend to winding trajectories~\cite{Kim:2004rp} in the field space of axions. A particularly simple case of such trajectories arises in the presence of two axions upon Higgsing~\cite{Dvali:2005an,Kaloper:2008fb} a linear combination of them~\cite{Hebecker:2015rya}. It has been argued more generally in~\cite{Saraswat:2016eaz} that such theories, even if they satisfy the WGC in their Coulomb phase, could effectively violate it (at least in some of its forms) in their Higgs phase. We would like to study next how the physical setups and constraints described above behave under Higgsing in a system with multiple $B_2$ forms.

Let us consider a particular system of two 2-form gauge fields $B_{i}$, with  $i=1,2$, coupled to a single one-form field $A$ in a Stueckelberg-like manner
\be\label{eq:mixedlag}
S=\int d^4x\sqrt{-g}\left[-\frac{1}{f^2} (|H_1|^2+|H_2|^2) -\frac{1}{4g^2}|dA+B_1+NB_2|^2 \right]\,.
\ee
Here $N$ is a (large) integer, $H_i=dB_i$ and, for simplicity, we have taken identical axion decay constants $f_1=f_2\equiv f$. The coupling between $A$ and $B_i$ gives a mass to the linear combination $B_a\propto B_1+N B_2$. The orthogonal combination $B_b\propto-NB_1+B_2$ remains (perturbatively) massless. It will feature the desired small coupling $f_{eff}\sim f/N$ and can hence be used to construct our axionic BHs. We can rewrite~\eqref{eq:mixedlag} in terms of these fields as 
\be\label{eq:mixedlag2}
S_{A,B_a}=\int d^4x\sqrt{-g}\left[-\frac{1+N^2}{f^2} (|H_a|^2+|H_b|^2) -\frac{1}{4g^2}|dA+(1+N^2)B_a|^2 \right]
\ee
where $B_1=B_a-N B_b$ and $B_2=N B_a+ B_b$. 

The Lagrangian~\eqref{eq:mixedlag} is invariant under the gauge transformations: $\{B_1,\,B_2,\, dA\}\to \{ B_1 + d\Lambda_1,\,B_2+d\Lambda_2,\,  dA-d\Lambda_1-Nd\Lambda_2\}$. For a spacetime with a non-trivial two-cycle $\sigma$, we can define the variables
\be
a\equiv \frac{1}{2\pi} \int_\sigma dA\,,\qquad b_i=\frac{1}{2\pi}\int_\sigma B_i\,.
\ee
Under gauge transformations with $\int_\sigma d\Lambda_i=2\pi c_i$, these variables shift as $\{b_1,\,b_2,\,a\}\to\{b_1+c_1,\,b_2 +c_2,\,a-c_1-Nc_2\}$. These continuous shift symmetries are of course broken to discrete periodicities by the presence of strings.

As mentioned before, the axionic hair can only be measured if there exist strings in the spectrum charged under the $B_i$. We will assume the existence of strings $\Sigma_1$ and $\Sigma_2$ coupled to $B_1$ and $B_2$ with unit charge, i.e.
\be\label{eq:spectrum}
S_{str}=\int_{\Sigma_1} B_1+\int_{\Sigma_2} B_2\,.
\ee
This spectrum determines the periodicity of the ``Wilson line" type variables $b_i$. The path integral measure includes $e^{iS_{str}}$ (in turn the phase measured by Aharonov-Bohm interference experiments) which is only invariant under large gauge transformations with $c_i\in \mathbb{Z}$. Hence, the continuous shift symmetries are broken to discrete periodicities:
\be\label{eq:shifts}
\{b_1,\, b_2,\,a\}\to \{b_1+1,\,b_2,\,a-1\}\,, ~~ \text{and} ~~ \{b_1,\,b_2,\,a\}\to \{b_1,\,b_2+1,\,a-N\}\,.
\ee

The normalization of $B_a$ and $B_b$ in~\eqref{eq:mixedlag2} has been chosen so that strings still carry integral charges, i.e.
\be\label{eq:spectrum2}
S_{str}=\int_{\Sigma_1} (B_a-NB_b)+\int_{\Sigma_2} (NB_a+B_b)\,.
\ee
A convenient basis for the space of large gauge transformations~\eqref{eq:shifts} is given by
\be\label{eq:shifts2}
\{b_a,\, b_b,\,a\} \to \{b_a,\,b_b+1,\,a\}\,, ~~ \text{and} ~~ \{b_a,\,b_b,\,a\}\to \{b_a-\frac{1}{1+N^2},\,b_b+\frac{N}{1+N^2},\,a+1\}\,,
\ee
where $b_a$ and $b_b$ are defined analogously to $b_1$ and $b_2$. We see that $b_b$, the massless field, still has unit periodicity.

We have constructed an effective theory of a single light two-form $B_b$ with small coupling $f_{eff}$. Following the arguments of previous sections, we would naively conclude that the strings in this setup should be extremely light, otherwise an inconsistency would arise. As we discuss next, however, there are extra ingredients in this theory that may allow for a different way out.

Consider a monopole of the theory~\eqref{eq:mixedlag2}, i.e. a particle that sources
\be
\frac{1}{2\pi}\int_{S^2} dA = a\,.
\ee
For concreteness, consider an (anti-) monopole of charge $a=-1$. Because of the Stueckelberg coupling between the $A$ and $B_a$, such a monopole carries a non-trivial value of $b_a$. From~\eqref{eq:mixedlag2} one can see that minimum action configurations satisfy $dA\to -(1+N^2) B_a$ as $r\to \infty$. This implies that, asymptotically, the magnetic monopole has
\be\label{eq:monopolecharge}
\left\{b_a=\frac{1}{1+N^2}\,,\,b_b=0\,,\,a=-1\right\}\,.
\ee
It looks as if the particle sources the `Wilson loop' variable associated with the heavy field $B_a$, but this is actually just a matter of gauge choice. From~\eqref{eq:shifts2}, we see that that~\eqref{eq:monopolecharge} can be equivalently written as
\be
\left\{b_a=0\,,\,b_b=\frac{N}{1+N^2}\,,\,a=0\right\}\,.
\ee
Given that the full periodicity of $b_b$ is unity, we see that the anti-monopole carries an axionic charge of about $1/N$ (of the maximum, which is of course the same as no axionic charge). Similar objects have been constructed in~\cite{Krauss:1990gx}.

\begin{figure}[ht]
\begin{center}
\includegraphics[width=8cm]{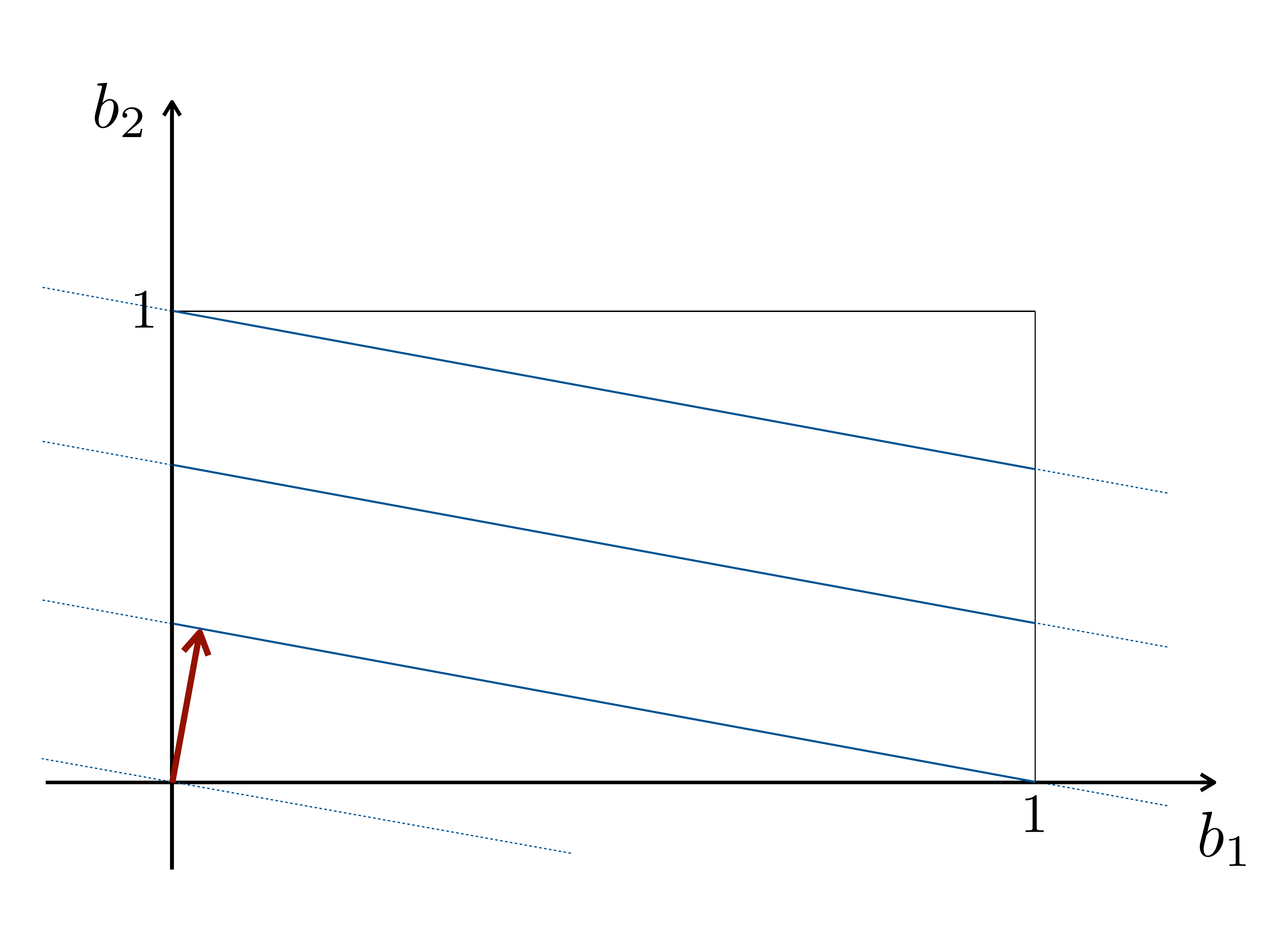}
\caption{Field space of $b_1$ and $b_2$. The subspace parametrized by the light axion for $N=3$ is shown. The arrow denotes the point associated with an $S^2$ loop around an anti-monopole.}\label{winding}
\end{center}
\end{figure}

This can be intuitively understood by looking at the $(b_1,b_2)$ field space. The monopole charge~\eqref{eq:monopolecharge} can be rewritten as 
\be
\left\{b_1=\frac{1}{1+N^2}\,,\,b_2=\frac{N}{1+N^2}\,,\,a=-1\right\}\,.
\label{b1b2}
\ee
As should be clear from Fig.~\ref{winding}, the point in $(b_1,b_2)$ field space given by (\ref{b1b2}) can actually be viewed as a fractional value of the light $B_b$ `Wilson line'.

Crucially, this implies that, from the perspective of the low energy effective theory, a BH with appropriate axion charge (roughly a multiple of $1/N$, e.g. $K/N$ with $K\leq N/2$), can in principle decay. It would decay to $K$ anti-monopoles of the 1-form theory which was used to create the small axion decay constant $f_{eff}\simeq f/N$ in the first place. However, this clearly requires the relevant monopoles to be light enough. 

Let us consider an axionic BH with $b_b\approx 1/2$. If we denote the mass of the magnetic monopoles by $m$, these will start being emitted once the BH reaches a radius $R_m\sim 1/m$, which corresponds to a BH of mass $M_{bh}\sim M_P^2/m$. Since the BH is required to emit $~N/2$ of these monopoles before its decay, we should require that $M_{bh}\gtrsim N m/2$, that is, $m\lesssim M_P/\sqrt{N}$. 

This is still a conservative estimate, since  in principle both monopoles and anti-monopoles will be radiated. In the absence of axionic hair, the net axionic charge carried by these would be zero. The study of the dynamical effect that the axionic hair of the BH has in the process of (anti-)monopole emission, and how this leads to a discharge of the axionic BH is a complicated subject that we leave for future work. However, we expect that, once this effect is taken into account, the actual upper bound on the monopole masses will be significantly lower than $m\lesssim M_P/\sqrt{N}$.

To put our result in perspective, recall that in the un-Higgsed theory the existence of strings with tension $\sigma_{1,2}\sim \Lambda^2 \lesssim f M_P$ suffices to satisfy the WGC and to avoid the possible pathologies discussed in Sect.~\ref{sec:problems}. By contrast, the situation in the Higgs phase appears to be more involved. Indeed, at low energies one encounters a theory with a single light field $B_b$, with an effective coupling $f_{eff}\sim f/N$, while the strings are now much heavier than the expected cutoff $\Lambda_{eff}\lesssim \sqrt{ f_{eff}M_P}\sim \sqrt{f M_P/N}$. Our pathology of axionic BH evaporation appears to occur. At the same time, a solution specific to this type of construction suggests itself: The possible lightness of monopoles of the gauged 1-form theory cures the problem if $m\lesssim M_P/\sqrt{N}$. Notice in particular that if $f\sim M_P$, one concludes that these monopoles have to reside precisely at the expected effective cutoff scale $m\lesssim \Lambda_{eff}\sim M_P/\sqrt{N}$. One may be tempted to conclude that the mechanisms of alignment and Higgsing are insufficient to generate an effective theory the cutoff of which evades the WGC: New low-scale physics may be forced to enter in a different sector.

Clearly, other conclusions are also possible. First of all, it may simply be impossible to construct the theory with $A$, $B_1$ and $B_2$ and the required large $N$ in consistent quantum gravity models, even though the WGC does not directly forbid such a situation. Furthermore, it is conceivable that such models avoid our `axionic BH evaporation problem' in some other way, not related to light strings or light monopoles. Nevertheless, the possible way out through light monopoles looks intriguing.

\section{Conclusions}\label{sec:conclusions}

In this note we have presented a novel argument that suggests a parametric bound on the tension of strings in theories with 2-form fields at weak coupling. While the exact form of the bound depends on the precise rate of evaporation of small axionic black holes, under certain assumptions we recover the constraints expected from the WGC. 
Our arguments are not air-tight, and allow for ways out that do not involve light strings. If the evaporation of black holes slowed down at late stages  (with respect to the standard rate of Hawking radiation), or if for some reason the semiclassical description of the evaporation became invalid at a stage earlier than expected, our bounds would become weaker. In particular, a long-lived remnant black hole would by itself resolve the apparent contradiction. In our opinion, however, the most natural solution relies on the existence of light strings.

We have further applied our arguments to a setup in which a light 2-form field with small effective decay constant $f_{eff}$ arises upon Higgsing an `aligned' linear combination of two 2-forms. Interestingly, rather than constraining the string spectrum, the evaporation of axionic black holes suggests the appearance of light states in a different sector, namely light monopoles which carry fractional amounts of axion hair. Given the prominent role played by similar models in the paradigm of large field (`natural') inflation, it would be interesting to understand whether our conclusions can be confirmed by alternative considerations and/or for objects of different dimensionality.

Concerning objects of different dimensionality, one encounters some obvious obstacles. First, in four or less flat space-time dimensions black holes are the only objects with horizon and a corresponding Hawking evaporation process. So our constraint on 2-forms does not generalize in any obvious way. One might be tempted to consider BTZ black holes in $d=3$ AdS and try to constrain the corresponding Wilson line and hence the 1-form gauge theory, but also here our logic does not go through straightforwardly. Crucially, we can not follow the evaporation of a macroscopic black hole all the way to empty space. Turning to space-time dimensions above four, we can in principle consider black branes in flat space, but then we  would face the Gregory-Laflamme instability. Other higher dimensional black objects such as black rings, whose horizons have richer topologies ($S^1\times S^{d-3}$), will also be interesting to consider. We leave the study of how to adopt our logic to these cases to future work.

Possibly, the right way to extend and generalize our argument is not via other black objects but via space-time topology changes. Indeed, the main idea we have used is the disappearance of an (effectively) non-trivial 2-cycle of our space time - the 2-cycle around the black hole horizon. Now, one might instead consider a geometry where two copies of $\mathbb{R}^d$ are linked by a throat with cross-section $S^p$ $(p\leq d-2)$, i.e., a wormhole, possibly fibered over an appropriate space of dimension $d-p-2$. Dynamically, this wormhole should collapse in pure gravity. But now, one may consider putting a non-zero $p$-form `Wilson-line' on its non-trivial cycle and demand a consistent dynamical transition to the trivial topology of two separate $\mathbb{R}^d$ spaces. It is conceivable that, following the logic of Sect.~\ref{sec:immediate}, one can recover the WGC bound. We leave a more detailed study of such possibilities to future work.

\section*{Acknowledgments}
We would like to thank William Cottrell and Miguel Montero for useful discussions. P.S. would like to thank Gary Shiu and the Hong Kong Institute for Advanced Study for hospitality during the completion of this work. This work was supported by the DFG Transregional Collaborative Research Centre TRR~33 ``The Dark Universe''.


\begin{thebibliography}{99}

\bibitem{ArkaniHamed:2006dz}
  N.~Arkani-Hamed, L.~Motl, A.~Nicolis and C.~Vafa,
  ``The String landscape, black holes and gravity as the weakest force,''
  JHEP {\bf 0706} (2007) 060 [hep-th/0601001].

\bibitem{Banks:2010zn} 
  T.~Banks and N.~Seiberg,
  ``Symmetries and Strings in Field Theory and Gravity,''
  Phys.\ Rev.\ D {\bf 83}, 084019 (2011)
  [arXiv:1011.5120 [hep-th]].

\bibitem{Gaiotto:2014kfa} 
  D.~Gaiotto, A.~Kapustin, N.~Seiberg and B.~Willett,
  ``Generalized Global Symmetries,''
  JHEP {\bf 1502}, 172 (2015)
  [arXiv:1412.5148 [hep-th]].

\bibitem{Banks:2006mm} 
  T.~Banks, M.~Johnson and A.~Shomer,
  ``A Note on Gauge Theories Coupled to Gravity,''
  JHEP {\bf 0609}, 049 (2006)
  [hep-th/0606277].
 
\bibitem{Cheung:2014vva}
  C.~Cheung and G.~N.~Remmen,
  ``Naturalness and the Weak Gravity Conjecture,''
  Phys.\ Rev.\ Lett.\  {\bf 113} (2014) 051601
  [arXiv:1402.2287 [hep-ph]].

\bibitem{Bellazzini:2015cra} 
  B.~Bellazzini, C.~Cheung and G.~N.~Remmen,
  ``Quantum Gravity Constraints from Unitarity and Analyticity,''
  Phys.\ Rev.\ D {\bf 93}, no. 6, 064076 (2016)
  [arXiv:1509.00851 [hep-th]].

\bibitem{Nakayama:2015hga} 
  Y.~Nakayama and Y.~Nomura,
  ``Weak gravity conjecture in the AdS/CFT correspondence,''
  Phys.\ Rev.\ D {\bf 92}, no. 12, 126006 (2015)
  [arXiv:1509.01647 [hep-th]].

\bibitem{Heidenreich:2015nta}
  B.~Heidenreich, M.~Reece and T.~Rudelius,
  ``Sharpening the Weak Gravity Conjecture with Dimensional Reduction,''
  JHEP {\bf 1602} (2016) 140
  [arXiv:1509.06374 [hep-th]].

\bibitem{Harlow:2015lma} 
  D.~Harlow,
  ``Wormholes, Emergent Gauge Fields, and the Weak Gravity Conjecture,''
  JHEP {\bf 1601}, 122 (2016)
  [arXiv:1510.07911 [hep-th]].
  
\bibitem{Ibanez:2015fcv}
  L.~E.~Ibanez, M.~Montero, A.~Uranga and I.~Valenzuela,
  ``Relaxion Monodromy and the Weak Gravity Conjecture,''
  JHEP {\bf 1604} (2016) 020
  [arXiv:1512.00025 [hep-th]].

\bibitem{Hebecker:2015zss}
  A.~Hebecker, F.~Rompineve and A.~Westphal,
  ``Axion Monodromy and the Weak Gravity Conjecture,''
  JHEP {\bf 1604} (2016) 157
  [arXiv:1512.03768 [hep-th]].

\bibitem{Heidenreich:2016aqi} 
  B.~Heidenreich, M.~Reece and T.~Rudelius,
  ``Evidence for a Lattice Weak Gravity Conjecture,''
  arXiv:1606.08437 [hep-th].
  
\bibitem{Montero:2016tif} 
  M.~Montero, G.~Shiu and P.~Soler,
  ``The Weak Gravity Conjecture in three dimensions,''
  JHEP {\bf 1610}, 159 (2016)
  [arXiv:1606.08438 [hep-th]].
  
  \bibitem{Saraswat:2016eaz}
  P.~Saraswat,
  ``The Weak Gravity Conjecture and Effective Field Theory,''\\
  arXiv:1608.06951 [hep-th].
  
\bibitem{Ooguri:2016pdq}
  H.~Ooguri and C.~Vafa,
  ``Non-supersymmetric AdS and the Swampland,''\\
  arXiv:1610.01533 [hep-th].
  
\bibitem{Freivogel:2016qwc} 
  B.~Freivogel and M.~Kleban,
  ``Vacua Morghulis,''
  arXiv:1610.04564 [hep-th].

\bibitem{Danielsson:2016mtx} 
  U.~Danielsson and G.~Dibitetto,
  ``The fate of stringy AdS vacua and the WGC,''
  arXiv:1611.01395 [hep-th].
  
\bibitem{Cottrell:2016bty}
  W.~Cottrell, G.~Shiu and P.~Soler,
  ``Weak Gravity Conjecture and Extremal Black Holes,''
  arXiv:1611.06270 [hep-th].

\bibitem{Banks:2016xpo} 
  T.~Banks,
  ``Note on a Paper by Ooguri and Vafa,''
  arXiv:1611.08953 [hep-th].
  
\bibitem{Hebecker:2017wsu} 
  A.~Hebecker, P.~Henkenjohann and L.~T.~Witkowski,
  ``What is the Magnetic Weak Gravity Conjecture for Axions?,''
  arXiv:1701.06553 [hep-th].

\bibitem{delaFuente:2014aca} 
  A.~de la Fuente, P.~Saraswat and R.~Sundrum,
  ``Natural Inflation and Quantum Gravity,''
  Phys.\ Rev.\ Lett.\  {\bf 114}, no. 15, 151303 (2015)
  [arXiv:1412.3457 [hep-th]].
  
\bibitem{Rudelius:2015xta}
  T.~Rudelius,
  ``Constraints on Axion Inflation from the Weak Gravity Conjecture,''
  JCAP {\bf 1509} (2015) no.09,  020
  [arXiv:1503.00795 [hep-th]].

\bibitem{Montero:2015ofa}
  M.~Montero, A.~M.~Uranga and I.~Valenzuela,
  ``Transplanckian axions!?,''
  JHEP {\bf 1508} (2015) 032 [arXiv:1503.03886 [hep-th]].

\bibitem{Brown:2015iha}
  J.~Brown, W.~Cottrell, G.~Shiu and P.~Soler,
  ``Fencing in the Swampland: Quantum Gravity Constraints on Large Field Inflation,''
  JHEP {\bf 1510} (2015) 023
  [arXiv:1503.04783 [hep-th]].

\bibitem{Bachlechner:2015qja} 
  T.~C.~Bachlechner, C.~Long and L.~McAllister,
  ``Planckian Axions and the Weak Gravity Conjecture,''
  JHEP {\bf 1601}, 091 (2016)
  [arXiv:1503.07853 [hep-th]].
  
\bibitem{Hebecker:2015rya}
  A.~Hebecker, P.~Mangat, F.~Rompineve and L.~T.~Witkowski,
  ``Winding out of the Swamp: Evading the Weak Gravity Conjecture with F-term Winding Inflation?,''
  Phys.\ Lett.\ B {\bf 748} (2015) 455
  [arXiv:1503.07912 [hep-th]].

\bibitem{Brown:2015lia} 
  J.~Brown, W.~Cottrell, G.~Shiu and P.~Soler,
  ``On Axionic Field Ranges, Loopholes and the Weak Gravity Conjecture,''
  JHEP {\bf 1604}, 017 (2016)
  [arXiv:1504.00659 [hep-th]].

\bibitem{Junghans:2015hba} 
  D.~Junghans,
  ``Large-Field Inflation with Multiple Axions and the Weak Gravity Conjecture,''
  JHEP {\bf 1602}, 128 (2016)
  [arXiv:1504.03566 [hep-th]].
   
   \bibitem{Heidenreich:2015wga}
  B.~Heidenreich, M.~Reece and T.~Rudelius,
  ``Weak Gravity Strongly Constrains Large-Field Axion Inflation,''
  JHEP {\bf 1512} (2015) 108 [arXiv:1506.03447 [hep-th]].

\bibitem{Kooner:2015rza} 
  K.~Kooner, S.~Parameswaran and I.~Zavala,
  ``Warping the Weak Gravity Conjecture,''
  Phys.\ Lett.\ B {\bf 759}, 402 (2016)
  [arXiv:1509.07049 [hep-th]].
  
\bibitem{Kaloper:2015jcz} 
  N.~Kaloper, M.~Kleban, A.~Lawrence and M.~S.~Sloth,
  ``Large Field Inflation and Gravitational Entropy,''
  Phys.\ Rev.\ D {\bf 93}, no. 4, 043510 (2016)
  [arXiv:1511.05119 [hep-th]].
  
\bibitem{Kappl:2015esy} 
  R.~Kappl, H.~P.~Nilles and M.~W.~Winkler,
  ``Modulated Natural Inflation,''
  Phys.\ Lett.\ B {\bf 753}, 653 (2016)
  [arXiv:1511.05560 [hep-th]].
  
\bibitem{Choi:2015aem} 
  K.~Choi and H.~Kim,
  ``Aligned natural inflation with modulations,''
  Phys.\ Lett.\ B {\bf 759}, 520 (2016)
  [arXiv:1511.07201 [hep-th]].
  
\bibitem{Hebecker:2016dsw} 
  A.~Hebecker, P.~Mangat, S.~Theisen and L.~T.~Witkowski,
  ``Can Gravitational Instantons Really Constrain Axion Inflation?,''
  arXiv:1607.06814 [hep-th].

\bibitem{Klaewer:2016kiy} 
  D.~Klaewer and E.~Palti,
  ``Super-Planckian Spatial Field Variations and Quantum Gravity,''
  arXiv:1610.00010 [hep-th].
  
\bibitem{Herraez:2016dxn} 
  A.~Herraez and L.~E.~Ibanez,
  ``An Axion-induced SM/MSSM Higgs Landscape and the Weak Gravity Conjecture,''
  arXiv:1610.08836 [hep-th].

\bibitem{Dolan:2017vmn} 
  M.~J.~Dolan, P.~Draper, J.~Kozaczuk and H.~Patel,
  ``Transplanckian Censorship and Global Cosmic Strings,''
  arXiv:1701.05572 [hep-th].
  
\bibitem{Bowick:1988xh}
  M.~J.~Bowick, S.~B.~Giddings, J.~A.~Harvey, G.~T.~Horowitz and A.~Strominger,
  ``Axionic Black Holes and a Bohm-Aharonov Effect for Strings,''
  Phys.\ Rev.\ Lett.\  {\bf 61} (1988) 2823.

\bibitem{Dvali:2016mur} 
  G.~Dvali and A.~Gußmann,
  ``Skyrmion Black Hole Hair: Conservation of Baryon Number by Black Holes and Observable Manifestations,''
  Nucl.\ Phys.\ B {\bf 913}, 1001 (2016)
  [arXiv:1605.00543 [hep-th]].

\bibitem{Dvali:2016sac} 
  G.~Dvali and A.~Gußmann,
  ``Aharonov-Bohm protection of black hole's baryon/ skyrmion hair,''
  arXiv:1611.09370 [hep-th].

\bibitem{Madrid}
  M.~Montero, A.~M.~Uranga and I.~Valenzuela,
  ``A Chern-Simons Pandemic,''
  arXiv:1702.06147 [hep-th].
  
\bibitem{Preskill:1990bm} 
  J.~Preskill and L.~M.~Krauss,
  ``Local Discrete Symmetry and Quantum Mechanical Hair,''
  Nucl.\ Phys.\ B {\bf 341}, 50 (1990).
  
\bibitem{Coleman:1991ku} 
  S.~R.~Coleman, J.~Preskill and F.~Wilczek,
  ``Quantum hair on black holes,''
  Nucl.\ Phys.\ B {\bf 378}, 175 (1992)
  [hep-th/9201059].
 
\bibitem{Krauss:1990gx} 
  L.~M.~Krauss and S.~J.~Rey,
  ``Duality, axion charge and quantum mechanical hair,''
  Phys.\ Lett.\ B {\bf 254}, 355 (1991).
  doi:10.1016/0370-2693(91)91168-U
  
 \bibitem{Cohen:1988sg} 
  A.~G.~Cohen and D.~B.~Kaplan,
  ``The Exact Metric About Global Cosmic Strings,''
  Phys.\ Lett.\ B {\bf 215}, 67 (1988).
  doi:10.1016/0370-2693(88)91072-6
  
\bibitem{Polchinski:2005bg} 
  J.~Polchinski,
  ``Open heterotic strings,''
  JHEP {\bf 0609}, 082 (2006)
  [hep-th/0510033].
 
\bibitem{Dvali:2005an} 
  G.~Dvali,
  ``Three-form gauging of axion symmetries and gravity,''
  hep-th/0507215.
  
\bibitem{Dvali:2005ws} 
  G.~Dvali, R.~Jackiw and S.~Y.~Pi,
  ``Topological mass generation in four dimensions,''
  Phys.\ Rev.\ Lett.\  {\bf 96}, 081602 (2006)
  [hep-th/0511175].
  
\bibitem{Dvali:2013cpa} 
  G.~Dvali, S.~Folkerts and A.~Franca,
  ``How neutrino protects the axion,''
  Phys.\ Rev.\ D {\bf 89}, no. 10, 105025 (2014)
  [arXiv:1312.7273 [hep-th]].
  
\bibitem{Dvali:2016uhn} 
  G.~Dvali and L.~Funcke,
  ``Small neutrino masses from gravitational $\theta$-term,''
  Phys.\ Rev.\ D {\bf 93}, no. 11, 113002 (2016)
  doi:10.1103/PhysRevD.93.113002
  [arXiv:1602.03191 [hep-ph]].
   
\bibitem{Garcia-Valdecasas:2016voz} 
  E.~Garc\'ia-Valdecasas and A.~Uranga,
  ``On the 3-form formulation of axion potentials from D-brane instantons,''
  arXiv:1605.08092 [hep-th].
  
\bibitem{Kim:2004rp}
  J.~E.~Kim, H.~P.~Nilles and M.~Peloso,
  ``Completing natural inflation,''
  JCAP {\bf 0501} (2005) 005
  [hep-ph/0409138].
  
  \bibitem{Kaloper:2008fb} 
  N.~Kaloper and L.~Sorbo,
  ``A Natural Framework for Chaotic Inflation,''
  Phys.\ Rev.\ Lett.\  {\bf 102}, 121301 (2009)
  [arXiv:0811.1989 [hep-th]].
   
\end{thebibliography}
\end{document}